\documentclass[conference]{IEEEtran}

\usepackage{graphicx}
\usepackage{subfigure,color}
\usepackage{tikz,comment}
\usetikzlibrary{calc}
\usetikzlibrary{intersections}

\DeclareTextSymbol{\degre}{OT1}{23}

\usepackage{supertabular}
\usepackage{empheq}

\newcommand{\be}{\begin{equation}}
\newcommand{\ee}{\end{equation}}
\newcommand{\bea}{\begin{eqnarray}}
\newcommand{\eea}{\end{eqnarray}}

\newcommand{\bbe}{\begin{empheq}[box=\fbox]{equation}}
\newcommand{\bbea}{\begin{empheq}[box=\fbox]{align}}

\newcommand{\rT}{r_{\text{Tx}}}
\newcommand{\rR}{r_{\text{Rx}}}

\newcommand{\cod}{_{\text{co}}}
\newcommand{\cad}{_{\text{ca}}}

\newcommand{\emid}{\ensuremath{_e}}
\newcommand{\recd}{\ensuremath{_r}}
\newcommand{\emiu}{\ensuremath{^e}}
\newcommand{\recu}{\ensuremath{^r}}
\newcommand{\LO}{\ensuremath{_\text{L.O.}}}
\newcommand{\beat}{\ensuremath{_\text{b}}}
\newcommand{\pps}{\ensuremath{_\text{pps}}}
\newcommand{\ppps}{\ensuremath{_\text{ppps}}}
\newcommand{\ini}{\ensuremath{_\text{ini}}}

\newcommand{\acc}{\ensuremath{^\text{acc}}}

\newcommand{\vxR}{\vec{x}_\text{Rx}}
\newcommand{\vxT}{\vec{x}_\text{Tx}}

\newcommand{\Dtc}{\Delta \tau_\text{mo}}

\newcommand{\Ol}{\mathcal{O}}

\usepackage{amsmath,amssymb}
\newcommand{\dd}{{\text{d}}}

\begin{document}

\title{Time and frequency transfer with a microwave link in the ACES/PHARAO mission}

\author{\authorblockN{P. Delva$^{*}$, F.
Meynadier, P. Wolf, C. Le~Poncin-Lafitte and P. Laurent}
\authorblockA{LNE-SYRTE, Observatoire de Paris, CNRS et UPMC,
\\61 avenue de l'Observatoire, 75014, Paris, France
\\$^{*}$Email: Pacome.Delva@obspm.fr}
}


%


\maketitle

\begin{abstract}

The Atomic Clocks Ensemble in Space (ACES/PHARAO mission), which will be
installed on board the International Space Station (ISS), uses a dedicated
two-way Micro-Wave Link (MWL) in order to compare the timescale generated on
board with those provided by many ground stations disseminated on the Earth.
Phase accuracy and stability of this long range link will have a key role in the
success of the ACES/PHARAO experiment. SYRTE laboratory is heavily involved in
the design and development of the data processing software : from theoretical
modelling and numerical simulations to the development of a software prototype.
Our team is working on a wide range of problems that need to be solved in order
to achieve high accuracy in (almost) real time. In this article we present some
key aspects of the measurement, as well as current status of the software's
development.

\end{abstract}


%
\IEEEpeerreviewmaketitle

\section{Introduction}

\begin{figure}[b]
\centering
\includegraphics[width=2.5in]{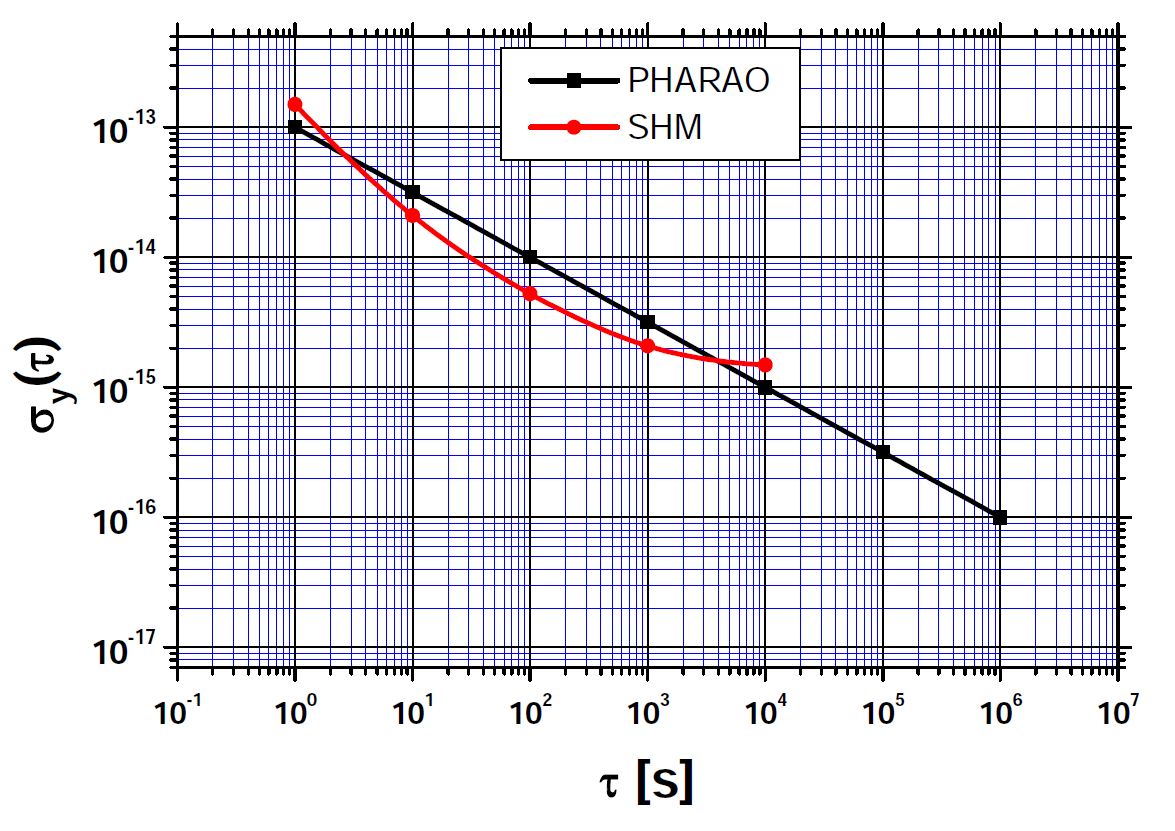}
\caption{ \label{f:sigma_y} PHARAO (Cesium clock) and SHM (hydrogen maser) expected performances in Allan
deviation.}
\end{figure}

\begin{figure}[b]
\centering
\includegraphics[width=2.5in]{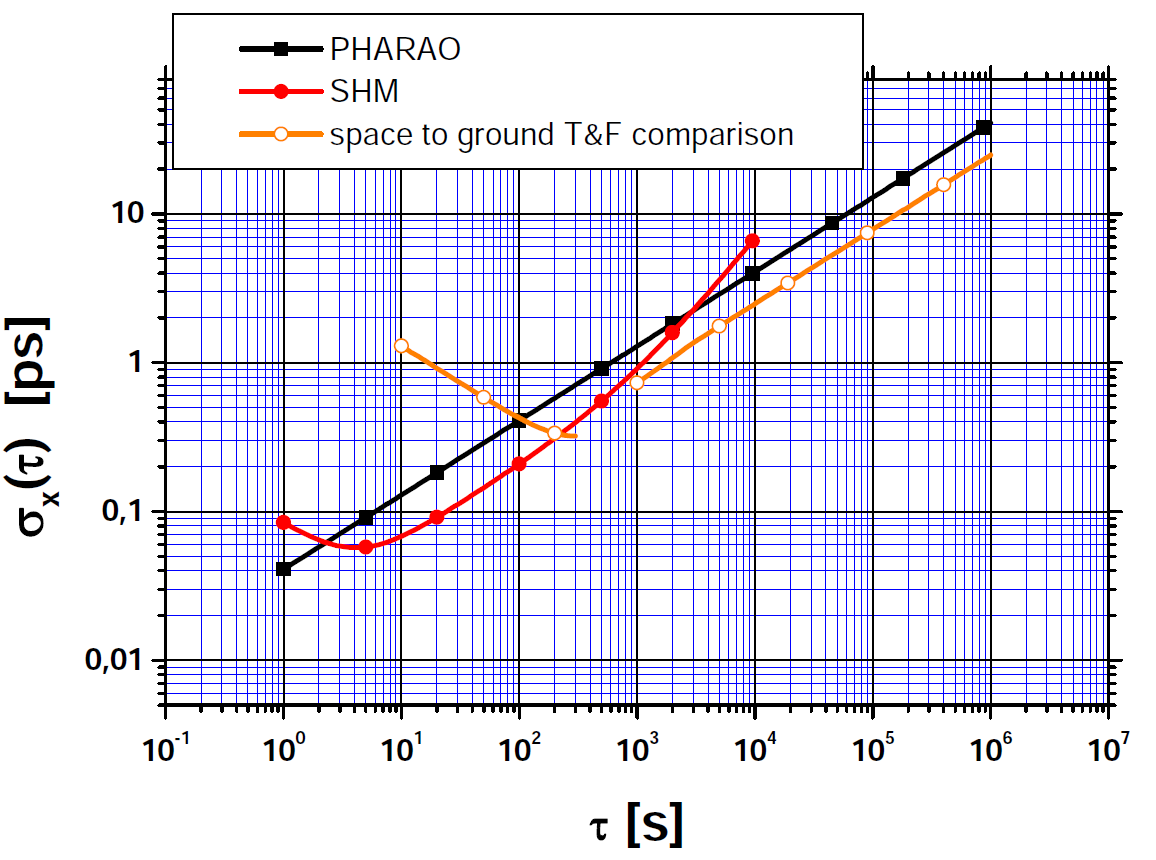}
\caption{\label{fig_second_case} Performance objective of the ACES clocks and
the ACES space-ground time and frequency transfer expressed in time
deviation.}
\label{f:sigma_x}
\end{figure}

The ACES/PHARAO mission is an international metrological space mission aiming at
realizing a time scale of high stability and accuracy on board the International
Space Station~(ISS). Relative frequency stability (ADEV) should be better
than $\sigma_y = 10^{-13} \cdot \tau^{-1/2}$, which corresponds to $3\cdot
10^{-16}$ after one day of integration (see fig.\ref{f:sigma_y}); time
deviation (TDEV) should be better than $2.1\cdot 10^{-14}\cdot \tau^{1/2}$,
which corresponds to 12~ps after one day of integration (see
fig.\ref{f:sigma_x}). Absolute frequency accuracy should be around
$10^{-16}$.


This mission is an international cooperation of more than 150 people. PI
laboratories are SYRTE/Paris Observatory, LKB/ENS and Neuch{\^a}tel Observatory,
and leading space agencies are the European Space Agency and CNES, the French space
agency. Many industries are involved, the main ones being EADS/Astrium, TimeTech
and Thales. All are working together to meet the scientific objectives of the
mission:
\begin{itemize}

\item Demonstrate the high permformance of the atomic clocks ensemble in the
space environment and the ability to achieve high stability on space-ground 
time and frequency transfer.

\item Compare ground clocks at high resolution on a world-wide basis using a 
link in the microwave domain. In common view mode, the link stability should
reach around 0.3~ps after 300~s of integration; in non-common view mode, it
should reach a stability of around 7~ps after 1~day of integration (see
fig.\ref{f:sigma_x}).

\item Perform equivalence principle tests. It will be possible to test
Local Lorentz Invariance and Local Position Invariance to unprecedent
accuracy by doing three types of tests: a test of gravitational red-shift,
drift of the fine structure constant and of anisotropy of light.

\end{itemize}

Besides these primary objectives, several secondary objectives
can be found in~\cite{LC-001}. For example, if the theory of general relativity
is considered as exact, then the measurement of gravitational redshifts
can be used to measure gravitational potential differences between
different clock locations. It is a new type of geodetic measurements using
clocks called relativistic geodesy.

In this article we describe in details the Micro-Wave Link (MWL) used in the
ACES/PHARAO mission, and developed by TimeTech (TT). First we describe one-way
and two-way links theoretically and introduce the SYRTE Team (ST) observables.
In the second part we describe how works the TT modem, and what is the link
between the modem observables (TT observables) and the ST observables. Finally
we present the status of the data analysis software and of the simulation we are
developing.

\section{The Micro-Wave Link (MWL)}

The Micro-Wave Link (MWL) will be used for space-ground time and frequency
transfer. A time transfer is the ability to synchronize distant clocks, i.e.
determine the difference of their displayed time for a given coordinate
time. The choice of time coordinate defines the notion of simultaneity,
which is only conventional. A frequency transfer is the ability to syntonize
distant clocks, i.e. determine the difference of clock frequencies for a
given coordinate time. Here we suppose that all clocks are perfect, i.e. their
displayed time is exactly their proper time. Proper time $\tau$ is given in a
metric theory of gravity by relation: 
\bea
\label{pr_time}
c^2 \dd \tau^2 = - g_{\alpha \beta} \dd x^\alpha
\dd x^\beta ,
\eea

where $g_{\alpha \beta}$ is the metric, $c$ the velocity of light, $\{ x^\alpha
\}$ the coordinates and Einstein summation rule is used. We use in this article
the notation $[.]$, which is the coordinate~/~proper time transformation
obtained from eq.\eqref{pr_time}, and $T_{ij} = t_j-t_i$ for coordinate
time intervals\footnote{e.g. $[T_{12}]^A$ is the transformation of
coordinate time interval $T_{12}$ in proper time of clock $A$, and $[\Delta
\tau^A]^t$ is the transformation of proper time interval of clock $A$ in
coordinate time $t$.}.

The MWL is composed of three signals of different frequencies: one uplink at
frequency $\simeq 13.5$~GHz, and two downlinks at $\simeq 14.7$~GHz and 2.2~GHz.
Measurements are done on the carrier itself and on a code which modulates the
carrier. The link is asynchronous: a configuration can be chosen by
interpolating observables. In the following we give a formal description of
one-way and two-way links for code observables. The principle for carrier
observables is the same except that periods cannot be identified, leading to a
phase ambiguity.

\subsection{One-way link}

\subsubsection{Experiment}

\begin{figure}[t]
\centerline{\subfigure[Sequence of events]{
\resizebox{\linewidth}{!} {
\input{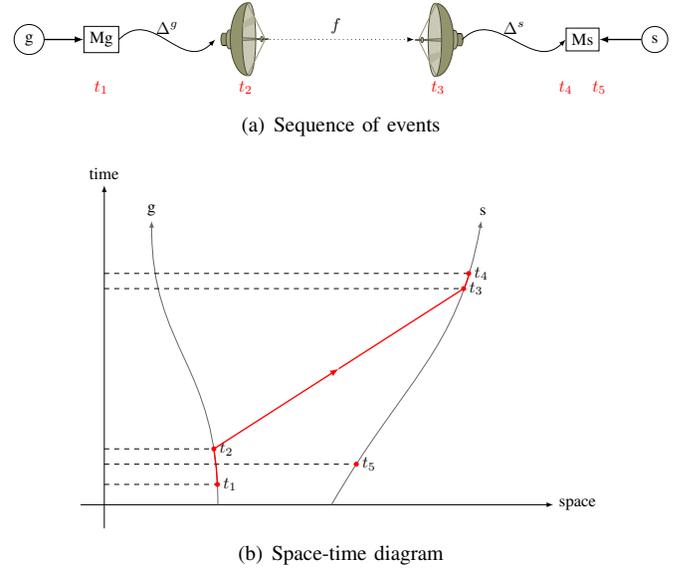}}
\label{f:1way1}}}
\centerline{\subfigure[Space-time diagram]{
\resizebox{0.8\linewidth}{!} {
\begin{tikzpicture}

\tikzstyle{signal} = [thick, color=red]
\tikzstyle{timeline} = [color=black!60]

\path (10,0) node (abs) {space};
\path (0,7) node (ord) {time};

\draw[-latex] (-0.5,0) -- (abs);
\draw[-latex] (0,-0.5) -- (ord);


\path (2.4,0) coordinate (originA);
\path (4.8,0) coordinate (originB);

\path (1,6) node[above] (endA) {g};
\path (8,6) node[above] (endB) {s};

\draw[-latex,timeline] (originA) .. controls +(90:3cm) and +(-90:3cm) .. (endA)
    coordinate[pos=0.05] (t1) 
    coordinate[pos=0.15] (t2);

\draw[-latex,timeline] (originB) .. controls +(60:3cm) and +(-100:3cm) .. (endB)
    coordinate[pos=0.12] (t5)
    coordinate[pos=0.8] (t3)
    coordinate[pos=0.85] (t4);


\draw[dashed] let \p1 = (t1) in (0,\y1) -- (\p1) node[right] (t1l) {$t_1$};
\draw[dashed] let \p1 = (t2) in (0,\y1) -- (\p1) node[right] (t2l) {$t_2$};
\draw[dashed] let \p1 = (t3) in (0,\y1) -- (\p1) node[right] (t3l) {$t_3$};
\draw[dashed] let \p1 = (t4) in (0,\y1) -- (\p1) node[right] (t4l) {$t_4$};
\draw[dashed] let \p1 = (t5) in (0,\y1) -- (\p1) node[right] (t5l) {$t_5$};


\foreach \e in {t1,t2,t3,t4,t5} {
  \fill[red] (\e) circle(1.5pt);
}


\begin{scope}
    \clip (t1) rectangle (t2);
    \draw[-latex, signal] (originA) .. controls +(90:3cm) and +(-90:3cm) .. (endA);
\end{scope}

\coordinate (tmp) at ($(t2)!0.5!(t3)$);
\draw[-latex, signal] (t2) -- (tmp);
\draw[signal] (tmp) -- (t3);

\begin{scope}
    \clip (t3) rectangle (t4);
    \draw[-latex, signal] (originB) .. controls +(60:3cm) and +(-100:3cm) ..
    (endB);
\end{scope}

\end{tikzpicture}}
\label{f:1way2}
}}
\caption{Schematic representation of the one-way link.}
\label{f:1way}
\end{figure}

let's consider a one-way link between a ground and a space clock represented
respectively by subscript $g$ and $s$. The sequence of events is illustrated on
fig.\ref{f:1way}. At time coordinate $t_1$, clock $g$ displays time $\tau_1$
and modem Mg produces a code $C^1$. This code modulates a sinusoidal signal of
frequency $f$ and sent at coordinate time $t_2$ by antenna $g$. The delay
between the code production and its transmission by antenna $g$ is $\Delta^g =
[T_{12}]^g$, expressed in local frame of clock $g$. Antenna $s$ receives signal
$C^1$ at coordinate time $t_3$, and transmit it to modem Ms and clock $s$ which
receives it at coordinate time $t_4$, with a delay $\Delta^s = [T_{34}]^s$. Clock
$s$ displays time $\tau_1$ and modem Ms produces the code $C^1$ at coordinate
time $t_5$. 

We use superscript $g$ or $s$ on proper times $\tau$ for clocks $g$ or $s$, and
we express proper time as a function of coordinate time. Then we can write
\be
\label{e:1way1}
\tau_1 = \tau^g (t_1) = \tau^s (t_5) . 
\ee

We define the ST (SYRTE Team) observable $\Delta \tau^s$ given by modem Ms with: 
\be
\label{e:1way2}
\Delta \tau^s ( \tau^s (t_4) ) = \tau^s (t_5) - \tau^s (t_4) . 
\ee 
This observable is dated with proper time of clock $s$ when this
clock receives code $C^1$ from antenna $s$. It can be interpreted as the
difference between the time of production of code $C^1$ by clock $s$, and time
of reception of same code $C^1$ sent by clock $g$, all expressed in proper time of
clock $s$.

\subsubsection{Desynchronisation}
desynchronisation between clock $g$ and $s$ is written in an hypersurface
characterized by coordinate time $t=\text{constant}$. From
eqs.\eqref{e:1way1}-\eqref{e:1way2} it is straightforward to deduce
it for coordinate time~$t_4$:
\be
\label{e:desync}
\tau^s (t_4) - \tau^g (t_4) = - \Delta \tau^s \left( \tau^s (t_4) \right) - \left[ T_{23}
+ \left[ \Delta^g + \Delta^s \right]^t \right]^g
\ee
Similar formulas can be obtained for the desynchronization at coordinate times
$t_1$ and $t_5$. This expression has been obtained for the uplink, from ground
to space. To obtain the desynchronisation with downlink observables, all you
have to do is replace $g$ and $s$ in eq.\eqref{e:desync}.

\subsection{Two-way link}

\begin{figure}[t]
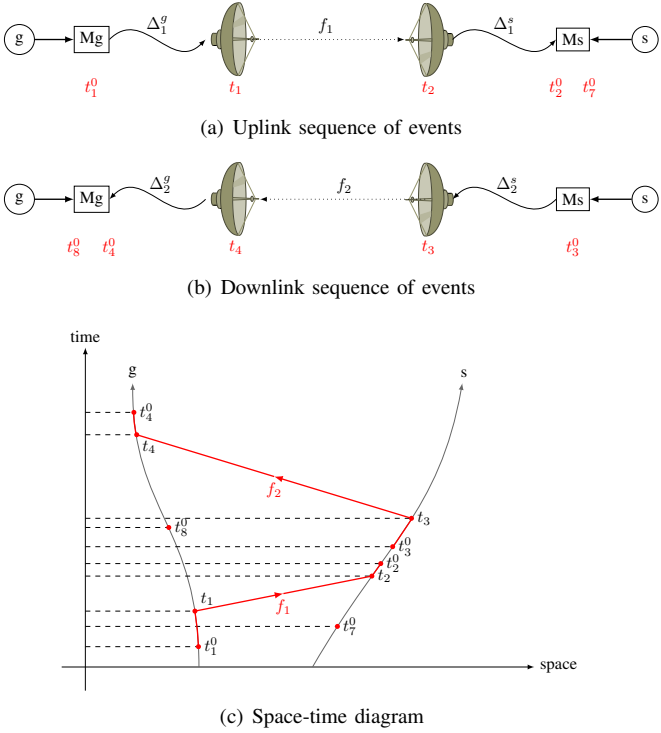

\subfigure[Uplink sequence of events]{
\resizebox{\linewidth}{!} {
\input{fig/twoway.tex}}
\label{f:2way1}}
\subfigure[Downlink sequence of events]{
\resizebox{\linewidth}{!} {
\input{fig/twoway2.tex}}
\label{f:2way2}}
\centerline{\subfigure[Space-time diagram]{
\resizebox{0.8\linewidth}{!} {
\begin{tikzpicture}

\tikzstyle{signal} = [thick, color=red]
\tikzstyle{timeline} = [color=black!60]

\path (10,0) node (abs) {space};
\path (0,7) node (ord) {time};

\draw[-latex] (-0.5,0) -- (abs);
\draw[-latex] (0,-0.5) -- (ord);


\path (2.4,0) coordinate (originA);
\path (4.8,0) coordinate (originB);

\path (1,6) node[above] (endA) {g};
\path (8,6) node[above] (endB) {s};

\draw[-latex,timeline] (originA) .. controls +(90:3cm) and +(-90:3cm) .. (endA)
    coordinate[pos=0.05] (t10) 
    coordinate[pos=0.15] (t1)
    coordinate[pos=0.47] (t80)
    coordinate[pos=0.85] (t4)
    coordinate[pos=0.92] (t40);

\draw[-latex,timeline] (originB) .. controls +(60:3cm) and +(-100:3cm) .. (endB)
    coordinate[pos=0.12] (t70)
    coordinate[pos=0.3] (t2)
    coordinate[pos=0.35] (t20)
    coordinate[pos=0.42] (t30)
    coordinate[pos=0.54] (t3);


\draw[dashed] let \p1 = (t10) in (0,\y1) -- (\p1) node[right] (t10l) {$t_1^0$};
\draw[dashed] let \p1 = (t1) in (0,\y1) -- (\p1) node[above right] (t1l) {$t_1$};
\draw[dashed] let \p1 = (t70) in (0,\y1) -- (\p1) node[right] (t70l) {$t_7^0$};
\draw[dashed] let \p1 = (t2) in (0,\y1) -- (\p1) node[right] (t2l) {$t_2$};
\draw[dashed] let \p1 = (t20) in (0,\y1) -- (\p1) node[right] (t20l) {$t_2^0$};
\draw[dashed] let \p1 = (t80) in (0,\y1) -- (\p1) node[right] (t80l) {$t_8^0$};
\draw[dashed] let \p1 = (t4) in (0,\y1) -- (\p1) node[below right] (t4l) {$t_4$};
\draw[dashed] let \p1 = (t40) in (0,\y1) -- (\p1) node[right] (t40l) {$t_4^0$};
\draw[dashed] let \p1 = (t30) in (0,\y1) -- (\p1) node[right] (t30l) {$t_3^0$};
\draw[dashed] let \p1 = (t3) in (0,\y1) -- (\p1) node[right] (t3l) {$t_3$};


\foreach \e in {t10,t1,t70,t2,t20,t80,t4,t40,t30,t3} {
  \fill[red] (\e) circle(1.5pt);
}


\begin{scope}
    \clip (t10) rectangle (t1);
    \draw[-latex, signal] (originA) .. controls +(90:3cm) and +(-90:3cm) .. (endA);
\end{scope}

\coordinate (tmp) at ($(t1)!0.5!(t2)$);
\draw[-latex, signal] (t1) -- (tmp);
\draw[signal] (tmp) -- (t2);
\path[signal] (tmp) node[below] (lf1) {$f_1$};

\begin{scope}
    \clip (t2) rectangle (t20);
    \draw[-latex, signal] (originB) .. controls +(60:3cm) and +(-100:3cm) ..
    (endB);
\end{scope}

\begin{scope}
    \clip (t4) rectangle (t40);
    \draw[-latex, signal] (originA) .. controls +(90:3cm) and +(-90:3cm) .. (endA);
\end{scope}

\coordinate (tmp) at ($(t3)!0.5!(t4)$);
\draw[-latex, signal] (t3) -- (tmp);
\draw[signal] (tmp) -- (t4);
\path[signal] (tmp) node[below] (lf2) {$f_2$};

\begin{scope}
    \clip (t30) rectangle (t3);
    \draw[-latex, signal] (originB) .. controls +(60:3cm) and +(-100:3cm) ..
    (endB);
\end{scope}

\end{tikzpicture}}
\label{f:2way2}
}}
\caption{Schematic representation of the two-way link.}
\label{f:2way}
\end{figure}

\subsubsection{Experiment}

let's consider now a link which is composed of two one-way links, between a
ground and a space clock represented respectively by subscript $g$ and $s$.  The
two sequences of events are illustrated on fig.~\ref{f:2way}. The uplink (from
ground to space) has a frequency $f_1$ and is represented by coordinate time
sequence $(t_1^0, t_1, t_2, t_2^0, t_7^0)$ (fig.\ref{f:2way1}). This link is
defined with the relation:
\be
\nonumber
\tau^g (t_1^0) = \tau^s (t_7^0) .
\ee
The downlink has a frequency $f_2$ and is represented by coordinate time
sequence $(t_3^0, t_3, t_4, t_4^0, t_8^0)$ (fig.\ref{f:1way2}). This link is
defined with the relation:
\be
\nonumber
\tau^s (t_3^0) = \tau^g (t_8^0) . 
\ee

\subsubsection{Desynchronisation in a two-way configuration}

a two-way configuration is defined by $\tau^g (t_1^0) = \tau^s (t_3^0)$, i.e.
the code $C^1$ of link $f_1$ is the same as code $C^2$ of link $f_2$, and they
are locally produced and sent at the same time: $t_1^0=t_8^0$ (at clock~$g$) and
$t_3^0=t_7^0$ (at clock~$s$).  Then we calculate desynchronisation between
clocks $g$ and $s$ at coordinate time $t^0_1$ as:
\be
\tau^s (t_1^0) - \tau^g (t_1^0) = \frac{1}{2} \left[ \left[
\Dtc^g (t_4^0) - \Dtc^s (t_2^0) \right]^t + T_{34} - T_{12} \right]^s
\ee
where we introduced the \emph{corrected} observables $\Dtc^g$ and $\Dtc^s$:
\be
\label{e:mod}
\begin{array}{lcl}
\Dtc^g (t_4^0) &=& \Delta \tau^g \left( \tau^g (t_4^0) \right) + \Delta^g_2 +
\Delta^s_2\\[0.1in]
\Dtc^s (t_2^0) &=& \Delta \tau^s \left( \tau^s (t_2^0) \right) + \Delta^g_1 +
\Delta^s_1.
\end{array}
\ee

\subsubsection{Desynchronisation in a $\Lambda$ configuration}

in the ACES/PHARAO mission we use the so-called $\Lambda$ configuration. This
configuration minimizes the error coming from the uncertainty on ISS
orbitography (in~\cite{duchayne2009} it has been shown that in this
configuration the requirement on ISS orbitography is around 10~m).  The
$\Lambda$ configuration is defined by $t_2=t_3$, i.e. code $C^2$ is sent at
antenna~$s$ when code $C^1$ is received at this antenna. This configuration is
obtained by interpolating the observables. Then it can be shown that 
desynchronisation between clocks $g$ and $s$ at coordinate time $t_2$ is:
\be
\label{eq:lambda}
\tau^s (t_2) - \tau^g (t_2) = \frac{1}{2} \left( \Dtc^g (t_4^0) - \Dtc^s (t_2^0)
 + \left[ T_{34} - T_{12} 
\right]^g \right) .
\ee

\subsection{Approximations}

\subsubsection{Coordinate~/~proper time transformation}
in equation~\eqref{eq:lambda} remains one
transformation from coordinate to proper time. We know that $T_{12} \sim T_{34}
\sim 1$~ms. During this time interval we can consider that the gravitational
potential and velocity of the ground station are constant. Therefore we can do
the approximation: 
\be \label{eq:approx} \left[ T_{34} - T_{12} \right]^g = ( 1 - \epsilon_g (t_2) )
\left( T_{34} - T_{12} \right) , \ee
where
\be 
\nonumber
\epsilon_g (t) = \dfrac{G M}{r_g(t) c^2} +
\dfrac{v_g^2(t)}{2 c^2} , \ee
$M$ is the Earth mass, $r_g(t)$ and $v_g(t)$ are the radial coordinate and the
coordinate velocity of the ground
clock at coordinate time $t$. 
Orders of magnitude of these corrective terms are:
\bea
\dfrac{G M}{r_g c^2} T_{34} &\sim& 0.6 \ \text{ps} \nonumber \\
\dfrac{v_g^2}{2 c^2} T_{34} &\sim& 0.002 \ \text{ps}\nonumber
\eea
The gravitational term is just at the limit of the required accuracy. The
velocity term is well below, so we can neglect it.
Final formula for desynchronisation is then:
\bea
\label{e:desyncfin}
\tau^s (t_2) - \tau^g (t_2) = \frac{1}{2} \left( \Dtc^g (t_4^0) - \Dtc^s
(t_2^0) \right. \nonumber\\
+ \left. \left( 1-\frac{G M}{r_g(t_2) c^2} \right) \left( T_{34} - T_{12} \right) \right)
.
\eea

\subsubsection{$\Lambda$ configuration}
in the $\Lambda$ configuration we suppose that $T_{23}=0$. However, this will
never be exactly 0 and it will be known with a precision $\delta T_{23}$.
This will add a supplementary delay $\delta (\tau^s - \tau^g)$ to
desynchronisation~\eqref{eq:lambda}:
\be
\nonumber
\delta (\tau^s - \tau^g) (t_2) = \left( \epsilon_g (t_2) - \epsilon_s
(t_2) \right) \delta T_{23}
\ee
Orders of magnitude are:
\bea
\dfrac{G M}{c^2} \left( \dfrac{1}{r_g} - \dfrac{1}{r_s} \right) &\sim& 2.8 \
\cdot 10^{-11} \nonumber\\
\dfrac{v_g^2-v_s^2}{2 c^2} &\sim& -3.3 \cdot 10^{-10} \nonumber
\eea
With the required accuracy on the MWL, $|\delta  (\tau^s - \tau^g)| \lesssim
0.3$~ps, we deduce the following constraint on $\delta T_{23}$:
\be \nonumber
\delta T_{23} \lesssim 
0.9 \ \text{ms}.
\ee
This constraint is much less constraining than the one coming from
orbitography, which is $\delta T_{23} \lesssim 1~\mu\text{s}$ (see~\cite{duchayne2009}).

\subsection{Atmospheric delays}

The downlink is composed of two one-way links of frequencies $f_2$ and $f_3$,
represented respectively by coordinate time sequence $(t_3^0, t_3, t_4,
t_4^0, t_8^0)$ and $(t_5^0, t_5, t_6, t_6^0, t_9^0)$. These two
links are affected by a ionospheric delay that depends on their respective
frequencies, whereas the tropospheric delay does not depend on the link
frequency (we neglect dispersive effects). We write:
\begin{align}
T_{34} & = \frac{R_{34}}{c} +
\Delta^\text{iono}_{34} (f_2) + \Delta^\text{tropo}_{34}\nonumber\\
&+ \dfrac{2GM}{c^3}
\ln \left( \dfrac{\rT(t_3)+\rR(t_4)+R_{34}}{\rT(t_3)+\rR(t_4)-R_{34}} \right) +
\Ol(c^{-4})
\label{eq:T34}\\
T_{56} & = \frac{R_{56}}{c} +
\Delta^\text{iono}_{56} (f_3) + \Delta^\text{tropo}_{56} \nonumber\\
&+ \dfrac{2GM}{c^3}
\ln \left( \dfrac{\rT(t_5)+\rR(t_6)+R_{56}}{\rT(t_5)+\rR(t_6)-R_{56}} \right) +
\Ol(c^{-4})
\label{eq:T56}
\end{align}
where $R_{ij} = |\vxR (t_j) - \vxT (t_i) |$ is the range, $\vxT$ and $\vxR$ are
respectively position vectors of space and ground antennas, $\rT =
|\vxT|$ and $\rR = |\vxR|$.

Ionospheric and tropospheric delays are around or below 100~ns, whereas
Shapiro delay (term in $c^{-3}$) is below 10~ps for the ACES/PHARAO mission
(see~\cite{duchayne:Thesis:2003} and fig.\ref{f:tof}).

\subsubsection{Ionospheric delay}

in order to deduce ionospheric delays, we combine the two ground
observables to be free of tropospheric delays. We obtain:
\be
\label{e:iono1}
\begin{array}{l}
\Delta \tau^g ( \tau^g(t^6_0) ) - \Delta \tau^g ( \tau^g (t^4_0) ) = \left[
T_{34} - T_{56} \right]^s \\
+ \left[ T_{46}^0 \right]^s - \left[ T_{46}^0 \right]^g + \Delta^s_2 -
\Delta^s_3 + \left[ \left[ \Delta^g_2 - \Delta^g_3 \right]^t \right]^s
\end{array}
\ee

Here we impose that $T_{46}^0 = 0$, ie. both signals are sent by clock~$s$ at
the same time. However, this will never be exactly zero, there will be a remaining
$\delta T_{46}^0$ introducing a timing error $\delta T \simeq  \left( \epsilon_s
(t_4^0) - \epsilon_g (t_4^0) \right) \delta T_{46}^0$.  With a required accuracy
$\delta T \lesssim 0.3$~ps, we obtain the following constraint:
\be
\nonumber
\delta T_{46}^0 \lesssim 0.9 \ \text{ms} .
\ee

We expect that $|T_{34} - T_{56}| \lesssim 100$~ns
(see~\cite{duchayne:Thesis:2003}); therefore we can neglect the coordinate to
proper time transformation in eq.\eqref{e:iono1}. We can also neglect this
transformation for the delays. Then eq.\eqref{e:iono1} is equivalent to:
\be
\label{e:iono3}
\Dtc^g (t_6^0) - \Dtc^g (t_4^0) = T_{34} - T_{56}
\ee
From eqs.\eqref{eq:T34}, \eqref{eq:T56} and~\eqref{e:iono3} we obtain:
\be
\label{eq:iono2}
\Delta^\text{iono}_{56} (f_3) - \Delta^\text{iono}_{34} (f_2) = \Dtc^g (t_4^0) -
\Dtc^g (t_6^0) + \dfrac{R_{34} - R_{56}}{c}
\ee
where we neglected the difference of the Shapiro delays between the two
downlinks, which can be shown to be completely negligable.

Now we can calculate $S$ the Slant Total Electron Content (STEC). The ionospheric delay
affects oppositely code and carrier and may be approximated as follows:
\begin{align}
\label{e:ionod1}
\Delta\cod^\text{iono} (f) &= \frac{40.308}{cf^2} S
+\frac{7527}{f^3}\int N_e \left( \vec{B} \cdot \vec{k} \right) \text{d} L\\
\Delta\cad^\text{iono} (f) &= -\frac{40.308}{cf^2} S
-\frac{7527}{2 f^3} \int N_e \left( \vec{B} \cdot
\vec{k} \right) \text{d}L \label{e:ionod2}
\end{align}
where $N_e$ is the local electron density along the path, STEC $S = \int N_e
\text{d}L$, $\vec{B}$ is the Earth's magnetic field and $\vec{k}$ the
unit vector along the direction of signal propagation. It has been shown that
higher order frequencies effect can be neglected for the determination of
desynchronisation~\cite{duchayne:Thesis:2003}.

We suppose that for a triplet of observables $\{ \Delta \tau^s (\tau^s(t_2^0)),
\Delta \tau^g (\tau^g(t_4^0)), \Delta \tau^g (\tau^g(t_6^0)) \}$, variations
of the direction of signal propagation and of magnetic field along
the line of sight does not change: $B \simeq B_0$. Then:
\begin{align}
\Delta\cod^\text{iono} (f) &= \frac{40.308}{cf^2} S
\left( 1 + \frac{7527 c}{40.308 f} B_0 \cos \theta_0 \right) \\
\Delta\cad^\text{iono} (f) &= - \frac{40.308}{cf^2} S
\left( 1 + \frac{7527 c}{80.616 f} B_0 \cos \theta_0 \right) , 
\end{align}
where $\theta_0$ is the angle between $\vec{B}$ and the direction of propagation
of signal $f_2$ and $f_3$. Then we obtain:
\be
\begin{array}{r}
\left[ \Delta^\text{iono}_{56} (f_3) - \Delta^\text{iono}_{34} (f_2)
\right]\cod = \dfrac{40.308}{c} \left( \dfrac{1}{f_3^2} -
\dfrac{1}{f_2^2} \right) S \\
\times \left[ 1 + \dfrac{7527 c}{40.308} \dfrac{f_2^3 -
f_3^3}{f_2 f_3 \left(f_2^2 - f_3^2 \right)} B_0 \cos \theta_0 \right]
\end{array}
\ee
\be
\begin{array}{r}
\left[ \Delta^\text{iono}_{56} (f_3) - \Delta^\text{iono}_{34} (f_2)
\right]\cad = - \dfrac{40.308}{c} \left( \dfrac{1}{f_3^2} -
\dfrac{1}{f_2^2} \right) S \\
\times \left[ 1 + \dfrac{7527 c}{80.616} \dfrac{f_2^3 -
f_3^3}{f_2 f_3 \left(f_2^2 - f_3^2 \right)} B_0 \cos \theta_0 \right]
\end{array}
\ee

These equations, together with equation~\eqref{eq:iono2}, give the STEC~$S$. The
value of $S$ can then be used to correct the uplink ionospheric delay.

\subsubsection{Tropospheric delay and range}

by adding ground and space observables of links $f_1$ and $f_2$ we obtain:
\be
\nonumber
\begin{array}{l}
\Delta \tau^s ( \tau^s(t_2^0) ) + \Delta \tau^g ( \tau^g(t_4^0) ) + \Delta^g_1 +
\Delta^g_2 \\[0.1in]
+ \left[ \left[ \Delta^s_1 + \Delta^s_2 \right]^t \right]^g = \left[ T_{23}^0
\right]^s - \left[ T_{23}^0 \right]^g - \left[ T_{12} + T_{34} \right]^g
\end{array}
\ee
As in the previous section, it can be shown that $\left[ T_{23}^0
\right]^s - \left[ T_{23}^0 \right]^g = 0$ if $T_{23}^0$ is known with a precision
$\delta T_{23}^0 \lesssim 0.9$~ms. We neglect the coordinate to proper time
transformations for delays and obtain:
\be
\nonumber
T_{12} + T_{34} = - \left( 1 + \dfrac{G M}{r_g (t_2) c^2} \right) \left(
\Dtc^s (t_2^0) + \Dtc^g (t_4^0) \right) ,
\ee
Then, neglecting Shapiro time delays we obtain:
\be
\nonumber
\begin{array}{r}
\dfrac{R_{12} + R_{34}}{c} = - \left( 1 + \dfrac{G M}{r_g (t_2) c^2} \right) \left(
\Dtc^s (t_2^0) + \Dtc^g (t_4^0) \right)\\[0.15in]
- \left( \Delta^{\text{iono}}_{12} (f_1) + \Delta^{\text{iono}}_{34} (f_2) 
+ \Delta^{\text{tropo}}_{12}  + \Delta^{\text{tropo}}_{34} \right)
\end{array}
\ee
This equation shows that range and tropospheric delays are degenerated.
Range can be calculated with a model for tropospheric delay, and tropospheric
delay can be calculated from an estimation of range.

\section{Micro-Wave Link modems}
\label{s:mwl}

\begin{figure*}[t]
\centering
\includegraphics[width=\linewidth]{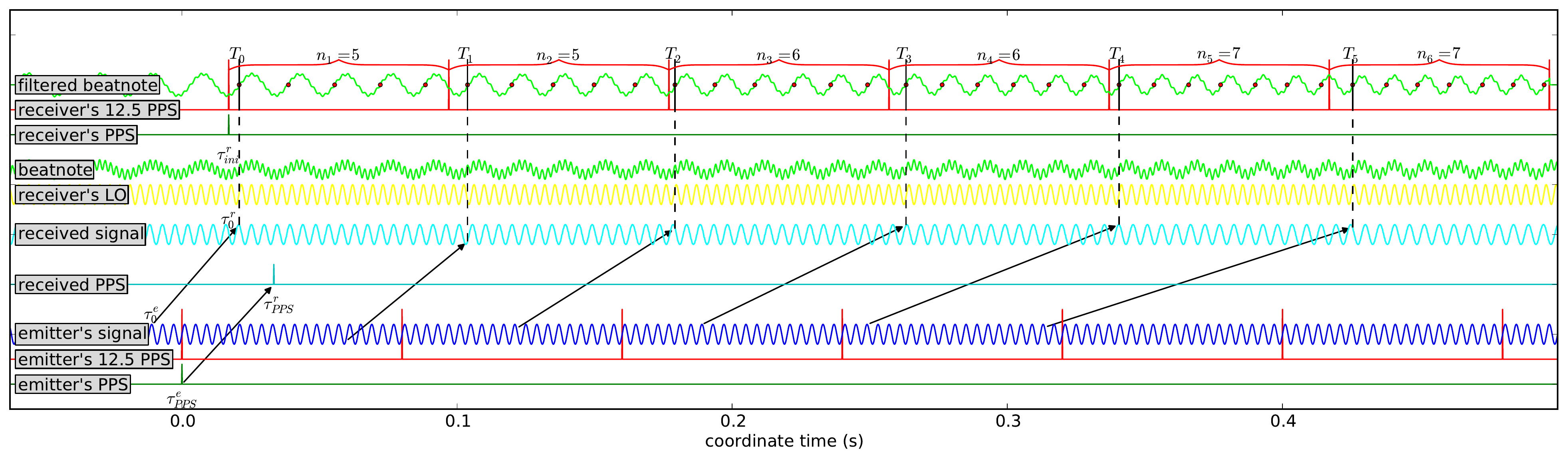}
\caption{ \label{f:Tm} Both emitter's and receiver's signals are represented against 
coordinate time scale. Red dots on the filtered beatnote indicate zero crossings
on ascending edge. $n_m$ is the number of red dots between two 12.5 PPS pulses.
$T_{m-1}$ is the proper time of the first red dot in the same sequence.
Propagation times are represented by black arrows.
Here signal frequency is much lower than in reality, and Doppler effect is strongly
magnified in order to show variation of $n_m$.}
\end{figure*}

We explain here basic principles of the ground/space modems developed by
TimeTech/Astrium for the ACES/PHARAO mission, that will be linked to the clocks
and the antennas. This principle is illustrated on fig.\ref{f:Tm}. At emitter
and receiver are generated a PPS signal (one Pulse Per Second), a 12.5~PPS (one
pulse every 80~ms, the period of measurements), and a periodic signal (either
code at 100~MHz or carrier). Let $e$ be the emitter and $r$ the receiver. A PPS
signal sent at local time $\tau\pps\emiu$ of the emitter is received at local
time $\tau\pps\recu$ of the receiver. Local time $\tau\pps\recu$ is recorded by
the modem for each received PPS.

When received, the periodic signal (blue) is mixed with a local oscillator
(yellow) which frequency is not far from the received frequency, and filtered to obtain
the low frequency part of the beatnote (green). The beatnote frequency is around 195~kHz
for code and 729~kHz for carrier. The receiver modem records the time of the
first ascending zero-phase of the beatnote signal after the 12.5~PPS signal.
We call this observable $T_m$, where $m$ is the number of the 80~ms
sequence. Finally, the modem counts the number of ascending zero-phase $n_m$
during sequence $m$.

$\tau\pps\recu$, $T_m$, $n_m$ and $m$ are the basic observables of the modem,
called TT observables, and are recorded for code and carrier signals. The modem
internal clock is reset every 4~s. However, code observables can be linked to
UTC time, which permits to solve the phase ambiguity between each
passage for code observables.

\subsection{From TT to ST observables}

\begin{figure}[b]
\centering
\resizebox{3in}{!} {
\begin{tikzpicture}

\tikzstyle{signal} = [thick, color=red]
\tikzstyle{timeline} = [color=black!60]
\tikzstyle{every node}=[font=\Large]
\path (10,0) node (abs) {space};
\path (0,7) node (ord) {time};

\draw[-latex] (-0.5,0) -- (abs);
\draw[-latex] (0,-0.5) -- (ord);


\path (2.4,0) coordinate (originE);
\path (4.8,0) coordinate (originR);

\path (1,6) node[above] (endE) {$e$};
\path (8,6) node[above] (endR) {$r$};

\draw[-latex,timeline, name path=Epath] (originE) .. controls +(90:3cm) and +(-90:3cm) .. (endE)
    coordinate[pos=0.15] (tau0E) 
    coordinate[pos=0.45] (tau1E);

\path[name path=signal0] (tau0E) -- +(20:10); 
\path[name path=signal1] (tau1E) -- +(20:10);

\draw[-latex,timeline, name path=Rpath] (originR) .. controls +(60:3cm) 
    and +(-100:3cm) .. (endR);




    \fill [name intersections={of=Rpath and signal0, by={tau0R}}]
      (tau0E) -- (tau0R);
    \fill [name intersections={of=Rpath and signal1, by={tau1R}}]
      (tau1E) -- (tau1R);


\foreach \e in {tau0E, tau1E, tau0R, tau1R} {
  \fill[red] (\e) circle(1.5pt);
}


%

\coordinate (tmp) at ($(tau0E)!0.5!(tau0R)$);
\draw[-latex, signal] (tau0E) -- (tmp);
\draw[signal] (tmp) -- (tau0R);

\coordinate (tmp) at ($(tau1E)!0.5!(tau1R)$);
\draw[-latex, signal] (tau1E) -- (tmp);
\draw[signal] (tmp) -- (tau1R);

\node[left] at (tau0E) {$\tau_0^e$};
\node[left] at (tau1E) {$\tau_1^e$};
\node[right] at (tau0R) {$\tau_0^r$};
\node[right] at (tau1R) {$\tau_1^r$};

\end{tikzpicture}}
\caption{ \label{f:signals} Link between phase, emitter and receiver local
times.}
\end{figure}
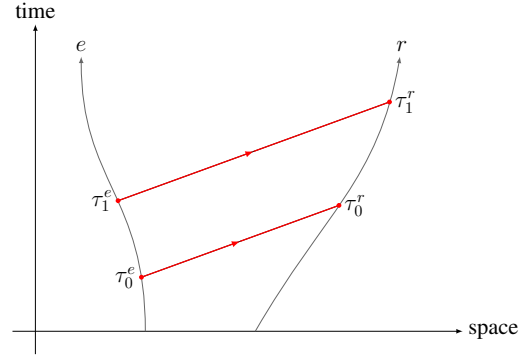

Let $\phi\emid (\tau\emiu)$ and $\phi\recd (\tau\recu)$ be respectively the
phase of emitted and received signals, changing with local time of emitter
and receiver. Let's consider two signals: one emitted at emitter local time
$\tau\emiu_0$ and received at receiver local time $\tau\recu_0$, and another
one, emitted at $\tau\emiu_1$ and received at $\tau\recu_1$ (see
fig.~\ref{f:signals}).
The phase increase between these two signals is equal at emitter and receiver:
\be
\label{e:dphi}
\phi\emid (\tau\emiu_1) - \phi\emid (\tau\emiu_0) = \phi\recd (\tau\recu_1) 
- \phi\recd (\tau\recu_0).
\ee
The received signal is mixed with a local oscillator signal such that the
beatnote phase is:
\be
\label{e:beat}
\phi\beat (\tau\recu) =
\left\{
    \begin{array}{ll}
    \phi\LO (\tau\recu) - \phi\recd (\tau\recu) & \text{(for code)}\\
    \phi\recd (\tau\recu) - \phi\LO (\tau\recu) & \text{\color{red}(for carrier)}
    \end{array}
\right.
\ee
Signs are different for code and carrier; we will write subsequent formulas in a
compact way with the sign in black for code and red for carrier. We assume
$\phi(\tau)= \omega \tau + \text{cst}$, where $\omega$ is the pulsation of the
considered signal. Then, from eqs.\eqref{e:dphi} and~\eqref{e:beat} we deduce:
\be
\label{e:tau}
\tau\emiu_1 - \tau\emiu_0 = \dfrac{\omega\LO}{\omega\emid} ( \tau\recu_1 -
\tau\recu_0 ) \stackrel{{\color{red} +}}{-} \dfrac{1}{\omega\emid} ( \phi\beat (\tau\recu_1) -
\phi\beat (\tau\recu_0) )
\ee
We introduce the ST observable, which links local time of emission to local time
of reception: $\Delta \tau\recu (\tau\recu) = \tau\emiu - \tau\recu$. From
eq.\eqref{e:tau} we get:
\be
\label{e:deltatau}
\begin{array}{ll}
\Delta\tau\recu_1 (\tau\recu_1) - \Delta\tau\recu_0 (\tau\recu_0) =& 
\left( \dfrac{\omega\LO}{\omega\emid} - 1 \right) ( \tau\recu_1 -
\tau\recu_0 ) \\[0.15in]
&\stackrel{{\color{red} +}}{-} \dfrac{1}{\omega\emid} ( \phi\beat (\tau\recu_1) -
\phi\beat (\tau\recu_0) )
\end{array}
\ee
Let's apply this formula to the MWL modem by introducing the TT obervables:
$\tau\recu_0 = T_{m-1}$ and $\tau\recu_1 = T_{m}$. From definition of $T_m$
and $n_m$ observables we know that $\phi\beat(T_m) - \phi\beat(T_{m-1}) = 2 \pi
n_m$. Then we deduce from eq.\eqref{e:deltatau}:
\be
\label{e:iter}
\begin{array}{ll}
\Delta\tau\recu_m (T_m) =& \Delta\tau\recu_{m-1} (T_{m-1})\\[0.15in]
& + \left(
\dfrac{\omega\LO}{\omega\emid} - 1 \right) ( T_m - T_{m-1} ) \\[0.15in]
& \stackrel{{\color{red} +}}{-} 
\dfrac{2\pi n_m}{\omega_e}
\end{array}
\ee

where $\Delta\tau\recu_m$ is the ST observable corresponding to sequence~$m$.
This recursive formula allows to find all ST observables from TT observables, if
the first term $\Delta\tau_0^r(T_0)$ is known. We notice than in case of zero
Doppler, the last two lines of the equation cancel, and the ST
observable $\Delta\tau\recu$ should be constant with local time.

Relative accuracy of ST observables during one passage of ISS is:
\be
\nonumber
\begin{array}{lcl}
\delta (\Delta\tau\recu_m) &\sim& \left| \dfrac{\omega\LO}{\omega\emid} - 1 \right|
\cdot \delta T_m\\[0.2in]
& \sim & \left\{
\begin{array}{ll}
\frac{195~\text{kHz}}{100~\text{MHz}}\cdot 10~\text{ns} \sim 20~ps & \text{(code)}\\[0.1in]
\frac{729~\text{kHz}}{13.5~\text{GHz}}\cdot 10~\text{ns} \sim 0.5~ps &
\text{(carrier)}
\end{array}
\right.
\end{array}
\ee
where accuracy of $T_m$ observables, $\delta T_m \sim 10$~ns, can be deduced from modem
internal clock, which frequency is around $100$~MHz. However, $\delta T_m$ is
underestimated here because other noise sources than the internal clock may
count. The goal here is not to do a precise accuracy budget but rather get a
lower limit.

\subsection{Initial term determination}

What is the first term of the iterative series~\eqref{e:iter}? Our goal is to
determine $\Delta\tau\recu_0 (T_0)$ with an absolute accuracy $<100$~ps,
which is required for ground-space time transfer. For frequency transfer, we
should be able to bridge the gap between two passages with an
accuracy depending on the duration between them, which can be read on
fig.\ref{f:sigma_x} (e.g. 1.5~ps for two passages separated by one orbital
period).

Let
\be
\label{e:tauini}
\Delta\tau\pps\recu (\tau\pps\recu) = \tau\pps\emiu - \tau\pps\recu =
\tau\ini\recu - \tau\pps\recu 
\ee
be the ST observables linked to TT observables $\tau\pps\recu$, and
$\tau\ini\recu$ be the receiver local time of generation of the PPS, which is by
definition of the experiment equal to the local time of emission of the same PPS
signal $\tau\pps\emiu$. As $\tau\pps\emiu$ can be guessed from a UTC tag,
$\Delta\tau\pps\recu (\tau\pps\recu)$ is known and can be linked to
$\Delta\tau\recu_0 (\tau\recu_0)$ with the help of eq.\eqref{e:deltatau}:
\be
\label{e:tau0}
\begin{array}{ll}
\Delta \tau_0\recu (\tau\recu_0) = & \Delta\tau\pps\recu (\tau\pps\recu) +
\left( \dfrac{\omega\LO}{\omega\emid} - 1 \right) \left( \tau_0\recu  -
\tau\pps\recu \right) \\[0.15in]
&- \dfrac{1}{\omega\emid} ( \Phi\beat (\tau_0\recu) - \Phi\beat
(\tau\pps\recu) )
\end{array}
\ee
From the values of $T_m$ and $n_m$ one can determine the beatnote phase at
$\tau\pps\recu$. The uncertainty in that determination is
$\delta\Phi_b(\tau\pps\recu) \sim \omega_b \cdot \delta T_m\sim 0.012$~rad.
The uncertainty of the two last terms on the rigth side of the equation is then $\sim
20$~ps, which is sufficient. However, $\Delta\tau\pps\recu$ is known with the
modem internal clock accuracy, i.e. $\delta( \Delta\tau\pps\recu ) \sim
10$~ns. Even averaging on a complete data set is not sufficient, reaching $\sim
580$~ps accuracy with 300 points. Then we need a method to obtain a precise PPS
observable.

\subsubsection{Precise PPS observable}

the emitter PPS is phase coherent with the code phase and the receiver PPS is
phase coherent with the local oscillator phase. Then we can use the internal counter
and the code phase observables to follow the phase precisely and derive a more
accurate pps observable that we call $\Delta\tau\ppps\recu$. 

The receiver local oscillator is phase coherent with the receiver PPS (it has
a zero crossing at $\tau\ini\recu$), then: \be
\label{eq:NLO} \Phi\LO (\tau\ini\recu) = 2\pi N\LO \ee where $2N\LO$ is an
integer. Similarly the emitted (and received) code is phase coherent with the
emitted (received) PPS, so we have \be \label{eq:NB} \Phi\emid (\tau\pps\emiu) =
\Phi\recd (\tau\pps\recu) = 2\pi N_B \ee where $N_B$ is an integer. Using
eqs.\eqref{e:beat}, \eqref{e:tauini}, \eqref{eq:NLO} and~\eqref{eq:NB} we
obtain:
\be 
\label{eq:ppps}
\Delta\tau\ppps\recu (\tau\pps\recu) = \dfrac{1}{\omega\LO} \left( 2 \pi N - \phi\beat 
(\tau\pps\recu) \right), 
\ee
where $2N=2(N\LO-N_B)$ is an integer. Provided we can determine the integer
$2N$ exactly, the uncertainty on $\Delta\tau\ppps\recu$ is $\delta
\phi\beat /\omega\LO \sim 20$~ps. The value of $N$ is determined by using the direct
measurement $\Delta\tau\pps\recu (\tau\pps\recu)$ in
eq.\eqref{eq:ppps} above:
\be
\label{eq:N}
N = \dfrac{1}{2\pi} \left( \omega\LO \Delta\tau\pps\recu (\tau\pps\recu) +
\Phi\beat (\tau\pps\recu) \right)
\ee
Uncertainty of the second term in~\eqref{eq:N} is $\sim 0.012/(2\pi) \sim
2\cdot 10^{-3}$, therefore negligible. However, uncertainty of the first term
is $\sim \omega\LO \cdot 10~\text{ns} / (2\pi) \sim 1$, which is insufficient. One solution is to
average over all PPS measurements of a continuous passage, which
should be sufficient for realistically useful passages (e.g. $> 50$~s). 

Finally using the precise PPS observable $\Delta\tau\ppps\recu(\tau\pps\recu)$
calculated from eq.\eqref{eq:ppps} in eq.\eqref{e:tau0} reaches the required
accuracy for time transfer.

\begin{figure*}
\centerline{\subfigure[Code observables]{\includegraphics[width=3in]{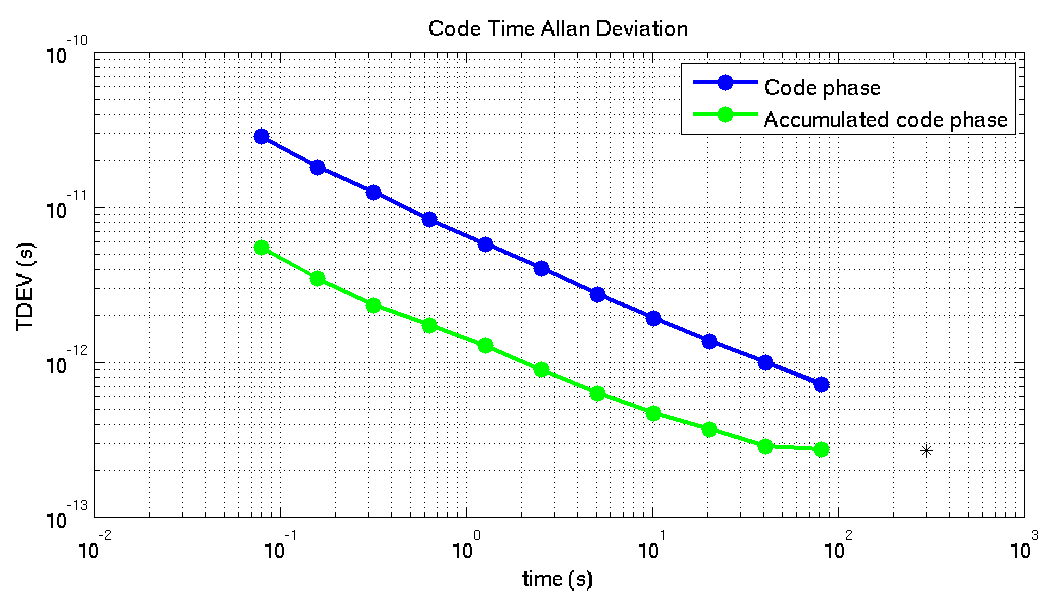}
\label{f:TTco}}
\hfil
\subfigure[Carrier observables]{\includegraphics[width=3in]{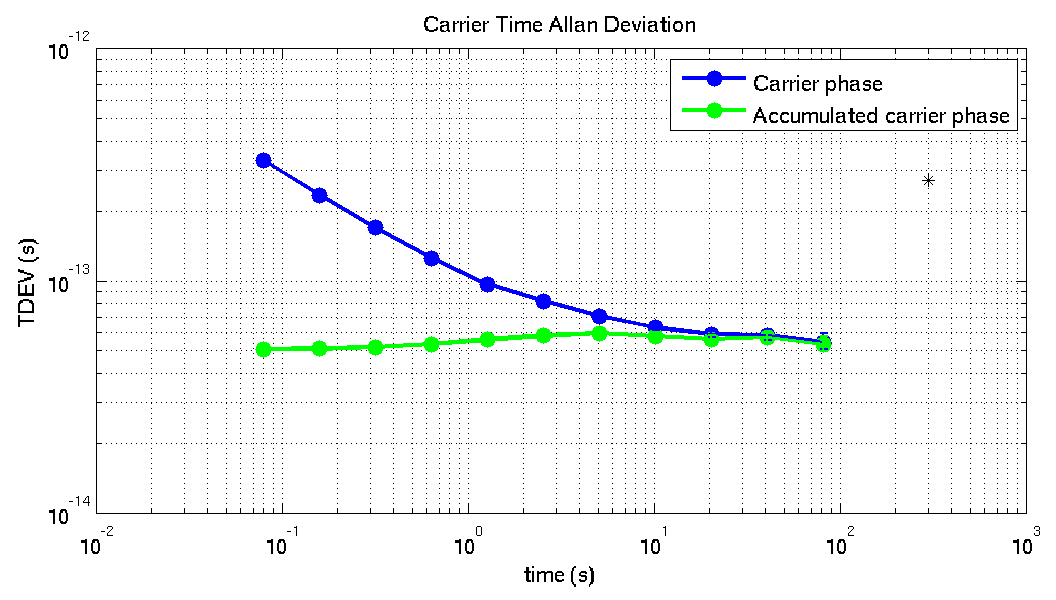}
\label{f:TTca}}}
\caption{Short term stability (TDEV) of ST observables, derived from either
code/carrier phase or accumulated code/carrier phase.}
\label{f:TT}
\end{figure*}

\subsubsection{Bridging the gap} to perform ground-space frequency comparisons on
long term one needs to link observables from two different ISS passages. It is then
necessary to determine precisely the time elapsed from one passage to another
thanks to the TT observables. Let's call $\Delta\tau_0(\tau_0)$ and
$\Delta\tau_0'(\tau'_0)$ the initial terms of two different passages (in this
section we omit the $r$ indices on ST observables $\Delta\tau\recu$ and on local
time $\tau\recu$). The error on their absolute determination is $\sim 20$~ps.
However we want to reach $\delta \left( \Delta\tau'_0(\tau'_0) -
\Delta\tau_0(\tau_0) \right) \lesssim x$, where $x$ is a specification that
depends on the duration separating the two passages and can be read from
fig.\ref{f:sigma_x} (e.g. 1.5~ps for one orbital period separation). 
From eq.\eqref{e:deltatau} we deduce:
\be
\begin{array}{ll}
\phi\beat (\tau'_0) - \phi\beat (\tau_0) = & -\omega\emid ( \Delta\tau'_0 (\tau'_0) 
- \Delta\tau_0 (\tau_0) )\\[0.1in]
&+ \left( \omega\LO - \omega\emid \right) (\tau'_0 - \tau_0 )
\end{array}
\label{eq:int1}
\ee
Moreover, we know that $\phi\beat (\tau'_0) - \phi\beat (\tau_0) = 2 \pi N_g$,
where $N_g$ is an unknown integer. We deduce that:
\be
\label{eq:Ng}
\begin{array}{ll}
N_g = \dfrac{1}{2\pi} &( -\omega\emid ( \Delta\tau'_0 (\tau'_0) - \Delta\tau_0
(\tau_0) ) \\
&+ \left( \omega\LO - \omega\recd \right) (\tau'_0 - \tau_0 ) )
\end{array}
\ee
It can be shown that the accuracies of the different terms in this equation are
sufficient to determine $N_g$ whithout ambiguity. Then we deduce:
\be
\label{eq:gap}
\Delta\tau'_0 (\tau'_0) - \Delta\tau_0 (\tau_0) = \left(
\dfrac{\omega\LO}{\omega\emid} - 1 \right) ( \tau'_0 - \tau_0 ) - \dfrac{2\pi N_g}{\omega\emid}
\ee
This equation links ST observables from two passages to $\sim$~20~ps, which is
not sufficient as can be seen from fig.\ref{f:sigma_x}. Using several pairs of
code observables to determine this quantity will not increase the accuracy, as
the error for each data pairs will be correlated. Two solutions can be
envisioned. One can use carrier phase observables. However, it remains to be
seen if the phase ambiguity (integer $N_g$) can be solved for carrier phase.
This will be studied in another article. Second solution would be to use
another observable from the MWL modem: the accumulated phase $T\acc_m$,
which is the sum of all dates of ascending zero-phase during sequence $m$.

In fig.\ref{f:TT} is shown the short-term stability of ST observables,
calculated either using $T_m$ or $T\acc_m$ observables, for code
(fig.\ref{f:TTco}) and carrier (fig.\ref{f:TTca}). It can be seen that using
$T\acc_m$ observables increases measurements accuracy. However, code accuracy
is still not sufficient to bridge the gap between two passages: we assume that
ST observables uncertainty is $\delta (\Delta\tau) \sim 3 \times \text{TDEV}_0$,
where $\text{TDEV}_0$ is the Time deviation for $\tau = 0.08$~s, the period of
measurements. Then, from fig.~\ref{f:TTco}, $\delta (\Delta\tau) \sim 90$~ps for
code phase and $\delta (\Delta\tau) \sim 18$~ps for accumulated code phase. This
is not sufficient to bridge any gap.  From fig.~\ref{f:TTca} we deduce $\delta
(\Delta\tau) \sim 0.9$~ps for carrier phase and $\delta (\Delta\tau) \sim
0.15$~ps for accumulated code phase. Any of these two observable is sufficient
to bridge a gap of 30~mn or larger. It remains to be seen how to solve the phase
ambiguity for carrier phase, which will be studied in another article.

\section{Simulation and data analysis}

\begin{figure}[b]
\centering
\includegraphics[width=0.8\linewidth]{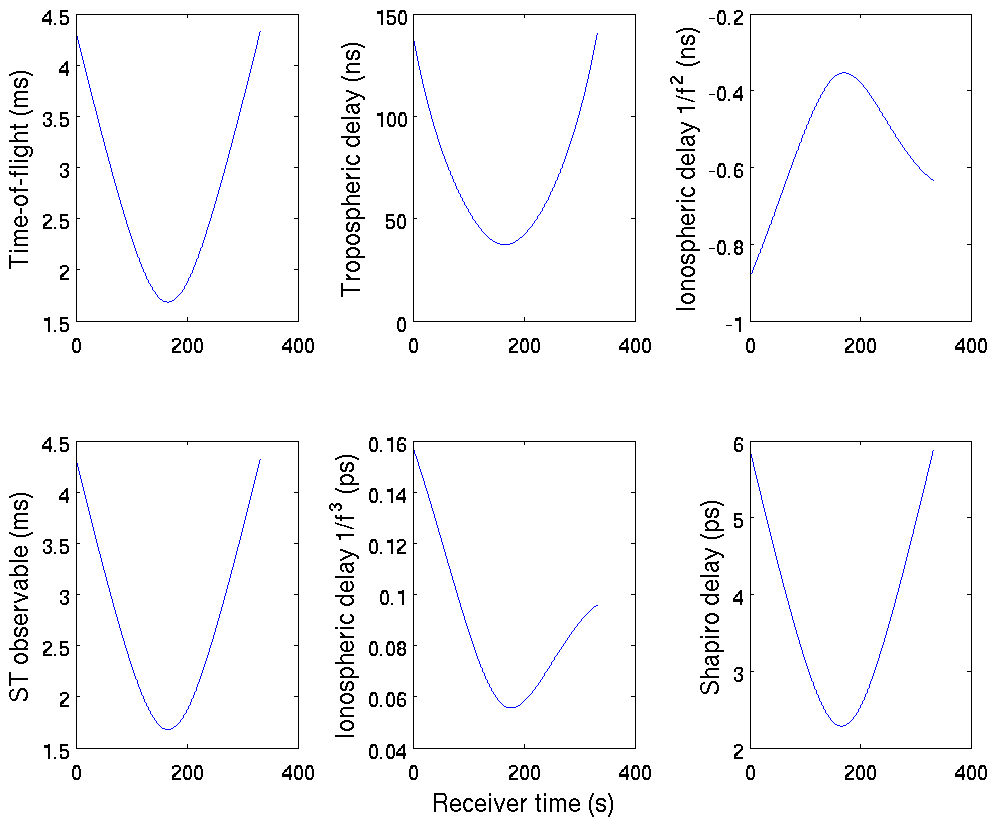}
\caption{Contribution of atmospheric and Shapiro delays in the simulated
time-of-fligth of the $f_1$ signal.}
\label{f:tof}
\end{figure}

\begin{figure}[t]
\centering
\includegraphics[width=1\linewidth]{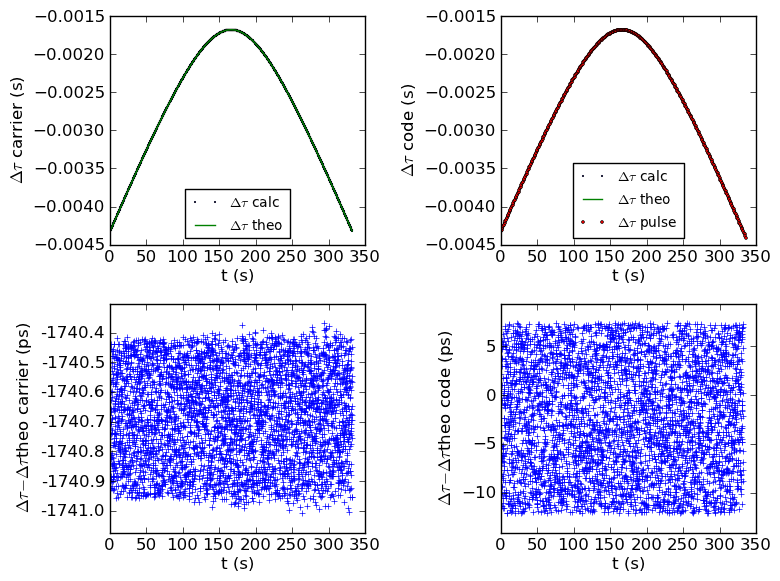}
\caption{Pre-processing software: comparison between ST
observables calculated from simulated TT observables, and theoretical ST
observables, for carrier (left) and code (right).}
\label{f:preproc}
\end{figure}

The SYRTE team writes an independant data analysis software, in order to make
the most of the ACES/PHARAO mission data. This software is written in Python
language. In order to test it, we wrote a simulation that generates (noisy) TT
observables, as well as theoretical ST observables. This simulation is written
in Matlab language, and is as much as possible independant from the data
analysis software.

\subsection{Simulation}

The simulation takes as input orbitography of the ISS and one Ground
Station~(GS) in a Celestial Reference System. From orbitography files it
simulates the proper times given by ISS and GS clocks, and the time transfer
between these two clocks, using a modelization of the MWL. Observables are given
in terms of TT and ST observables. Moreover, theoretical values of the scientific
products are given in order to test the data analysis software.

In fig.\ref{f:tof} we have plotted different contributions included in the
signal time-of-flight of signal $f_1\simeq 13.5$~GHz. Minimum elevation of ISS is
taken as 10\degre, and atmospheric parameters are temperature $T=298$~K,
pressure $p=1$~bar and water vapor pressure $e=0.5$~bar. A sinusoidal variation
is added to atmospheric parameters, with a period of 24~hours, and a Chapman
layer model is used to calculate the STEC. 

Tropospheric delay is dominant in the time-of-flight, with a value of several
10~ns. We used here a Saastamoinen model, which is not really reliable at low
elevation of ISS. The dispersive part of troposphere has not been taken into
account, and it remains to be seen if this is necessary. Ionosphere is
dispersive such taht the ionospheric delay can be separated in two
contributions: an effect that scales with $1/f^2$ and one that scales with
$1/f^3$ (see eqs.\eqref{e:ionod1}-\eqref{e:ionod2}). The second order
contribution is around 1~ns and the third contribution around 0.1~ps, below
mission accuracy. However these effects are much larger for the $f_3=2.25$~GHz
signal: around 40~ns for second order term and 20~ps for third order term. Then
third order terms cannot be ignored. Finally the Shapiro delay of several ps is
sligthly over the required accuracy.

\subsection{Data analysis}

An independant pre-processing software has been written, using equations from
sec.\ref{s:mwl}. It takes TT observables from the simulation, transform them to
ST observables and compare the result to the theoretical ST observables coming
from the simulation. One example can be seen on
fig.\ref{f:preproc}. It can be seen that ST observables are well recovered,
with a noise which is coherent with previous estimations. However here the noise
is underestimated because all noise sources has not be included (only the modem internal
clock noise). The absolute value of the ST code observable is found thanks to the
method of the initial term determination, with an uncertainty less than 20~ps
(explaining why the data cloud is not centered on 0). The phase ambiguity for
carrier observable has not been solved, explaining the constant bias between the
recovered and the theoretical ST carrier observables.

\begin{figure}[t]
\centering
\includegraphics[width=\linewidth]{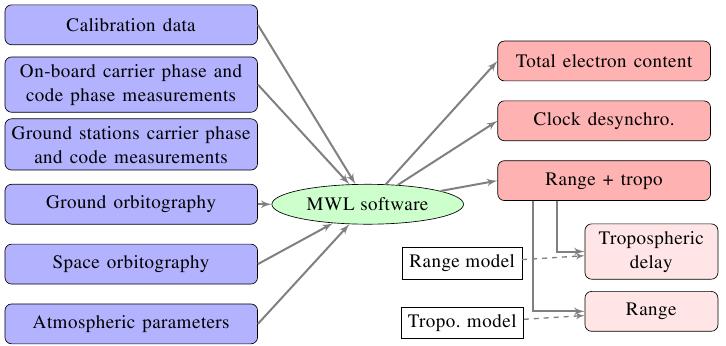}
\caption{Full data analysis software: illustration of inputs and outputs.}
\label{f:scheme}
\end{figure}

The full data analysis software is being written but not finished yet. We explain its
basic principle on fig.\ref{f:scheme}. A special care is taken for 
file naming, data classifying, file formats and conventions. Indeed many data
from several different sources will have to be used and these issues can be critical.
The software design has been build in a modular way, and now most of the
building blocks are written.

\section{Conclusion}

We have written a theoretical description of one-way and two-way satellite time
and frequency transfer and developed a model of the Micro-Wave Link in the frame
of the ACES/PHARAO mission. This description has been used to write a data
analysis software and a simulation to test it. The simulation is written in its
first version, and used to assess our pre-proceesing software. The design of the
data analysis software has been done in a modular way, and most of the building
blocks are ready.

Several questions remains: how to solve the phase ambiguity for the
carrier observable, what is the dispersive effect of the troposphere?

\end{document}